\documentclass[twocolumn,aps,pra,amsmath,amssymb,showpacs]{revtex4-1}
\usepackage[latin9]{inputenc}
\usepackage{color}
\usepackage{amsmath}
\usepackage{amssymb}
\usepackage{wasysym}
\usepackage{graphicx}
\usepackage{esint}
\usepackage[unicode=true,pdfusetitle,
 bookmarks=true,bookmarksnumbered=false,bookmarksopen=false,
 breaklinks=false,pdfborder={0 0 0},backref=false,colorlinks=true]
 {hyperref}

\makeatletter

\@ifundefined{textcolor}{}
{%
 \definecolor{BLACK}{gray}{0}
 \definecolor{WHITE}{gray}{1}
 \definecolor{RED}{rgb}{1,0,0}
 \definecolor{GREEN}{rgb}{0,1,0}
 \definecolor{BLUE}{rgb}{0,0,1}
 \definecolor{CYAN}{cmyk}{1,0,0,0}
 \definecolor{MAGENTA}{cmyk}{0,1,0,0}
 \definecolor{YELLOW}{cmyk}{0,0,1,0}
}

\usepackage{epsfig}\usepackage{amsfonts}

\usepackage[english]{babel}

\makeatother

\begin{document}

\title{Long-distance synchronization of unidirectionally cascaded optomechanical
systems}

\author{Tan Li, $^{1,2,3}$}

\author{Tian-Yi Bao, $^{2}$}

\author{Yan-Lei Zhang, $^{2,3}$ }

\author{Chang-Ling Zou, $^{2,3}$ }

\email{clzou321@ustc.edu.cn}

\author{Xu-Bo Zou, $^{2,3}$ }

\email{xbz@ustc.edu.cn}

\author{Guang-Can Guo, $^{2,3}$ }

\affiliation{$^{1}$ Zhengzhou Information Science and Technology Institute, Zhengzhou,
Henan 450004, China}

\affiliation{$^{2}$ Key Laboratory of Quantum Information, University of Science
and Technology of China, Hefei, Anhui 230026, China}

\affiliation{$^{3}$ Synergetic Innovation Center of Quantum Information and Quantum
Physics, University of Science and Technology of China, Hefei, Anhui
230026, China}
\begin{abstract}
Synchronization is of great scientific interest due to the abundant
applications in a wide range of systems. We propose a scheme to achieve
the controllable long-distance synchronization of two dissimilar optomechanical
systems, which are unidirectionally coupled through a fiber with light.
Synchronization, unsynchronization, and the dependence of the synchronization
on driving laser strength and intrinsic frequency mismatch are studied
based on the numerical simulation. Taking the fiber attenuation into
account, it's shown that two mechanical resonators can be synchronized
over a distance of tens of kilometers. In addition, we also analyze
the unidirectional synchronization of three optomechanical systems,
demonstrating the scalability of our scheme.
\end{abstract}


\maketitle

\section{\textit{\emph{Introduction}}}

Synchronization is a universal phenomenon in nature, where oscillators
with different intrinsic frequencies can adjust their rhythms to oscillate
in unison \cite{pikovsky2003synchronization,strogatz2003sync}. In
1660s, Huygens observed the synchronization of two pendulum clocks
hanging on a same wall \cite{nijhoff1893oeuvres}. Since then, synchronization
has been observed in a wide range of systems. For example, the coordination
of neurons \cite{izhikevich2007dynamical} and the regular flash of
glowworms colonies \cite{Buck22031968}. Synchronization is of importance
for both fundamental research and practical applications, since it
has the capacity to improve the precision \cite{PhysRevE.85.046214}
of frequency sources built from (electro)mechanical oscillators in
producing oscillating signals, which plays a critical role in time-keeping
\cite{Sivrikaya2004time}, sensing \cite{Bargatin2012Sensing} and
communication \cite{bregni2002synchronization}.

Synchronization has been demonstrated in many systems, such as Josephson
junctions \cite{PhysRevB.60.7575,PhysRevE.88.022908}, micro- and
nano- electromechanical systems \cite{PhysRevLett.111.084101,PhysRevB.87.144304,PhysRevLett.112.014101,PhysRevLett.114.034103},
ensembles of atoms \cite{PhysRevLett.113.154101}. Optomechanical
system (OMS) \cite{aspelmeyer2014cavityArticle,aspelmeyer2014cavityBook,kippenberg2008cavity}
is one of such platforms for synchronization research \cite{heinrich2011collective,Manipatruni2011Long,ying2014quantum,li2015criterion,shah2015synchronization},
and holds great potential for applications due to the easily fabrication,
high quality factor of optical resonators and strong optomechanical
coupling. The synchronization of OMSs have been predicted theoretically
\cite{PhysRevE.85.066203} and demonstrated in experiments \cite{zhang2012synchronization,bagheri2013photonic,zhang2015synchronization}.
For example, in two silicon nitride microdisks, spaced apart by 400nm,
two mechanical modes are synchronized by the coupling of two optical
modes \cite{zhang2012synchronization}. Two spatially separated 80
micrometers nanomechanical oscillators are also synchronized through
coupling to a same racetrack cavity \cite{bagheri2013photonic}.

However, those OMSs are coupled through local optical coupling between
cavities, while the greatest advantage of the light that can propagate
over very long distance is overlooked. Very recently, a long-distance
master-slave frequency locking has been realized between two OMSs
\cite{shah2015master}, while the light output from one OMS is converted
to radio frequency (RF) signal and the other OMS is injection locked
by using an electro-optic modulator (EOM) to modulate the input laser.
Extra elements required in this scheme, such as detectors and amplifiers
will introduce noises to such system and may limit the stability of
the system.

In this paper, we present a scheme to realize synchronization of cascaded
OMSs, where two OMSs are coupled through light propagating unidirectionally
in fiber, no extra detection of light is required. Through numerical
simulation, we observe the synchronization phenomenon and study the
influence of different systemic and external driving parameters on
synchronization. In practical applications in long distance synchronization,
we take the fiber attenuation into account, and confirm the synchronization
is possible for two OMSs over tens of kilometers. Last but not least,
we expand synchronization of two OMSs into synchronization of three
OMSs, which verifies the feasibility of unidirectional synchronization
of an OMSs array.

\section{\textit{\emph{Model}}}

\begin{figure}[th]
\begin{centering}
\includegraphics[width=8.5cm]{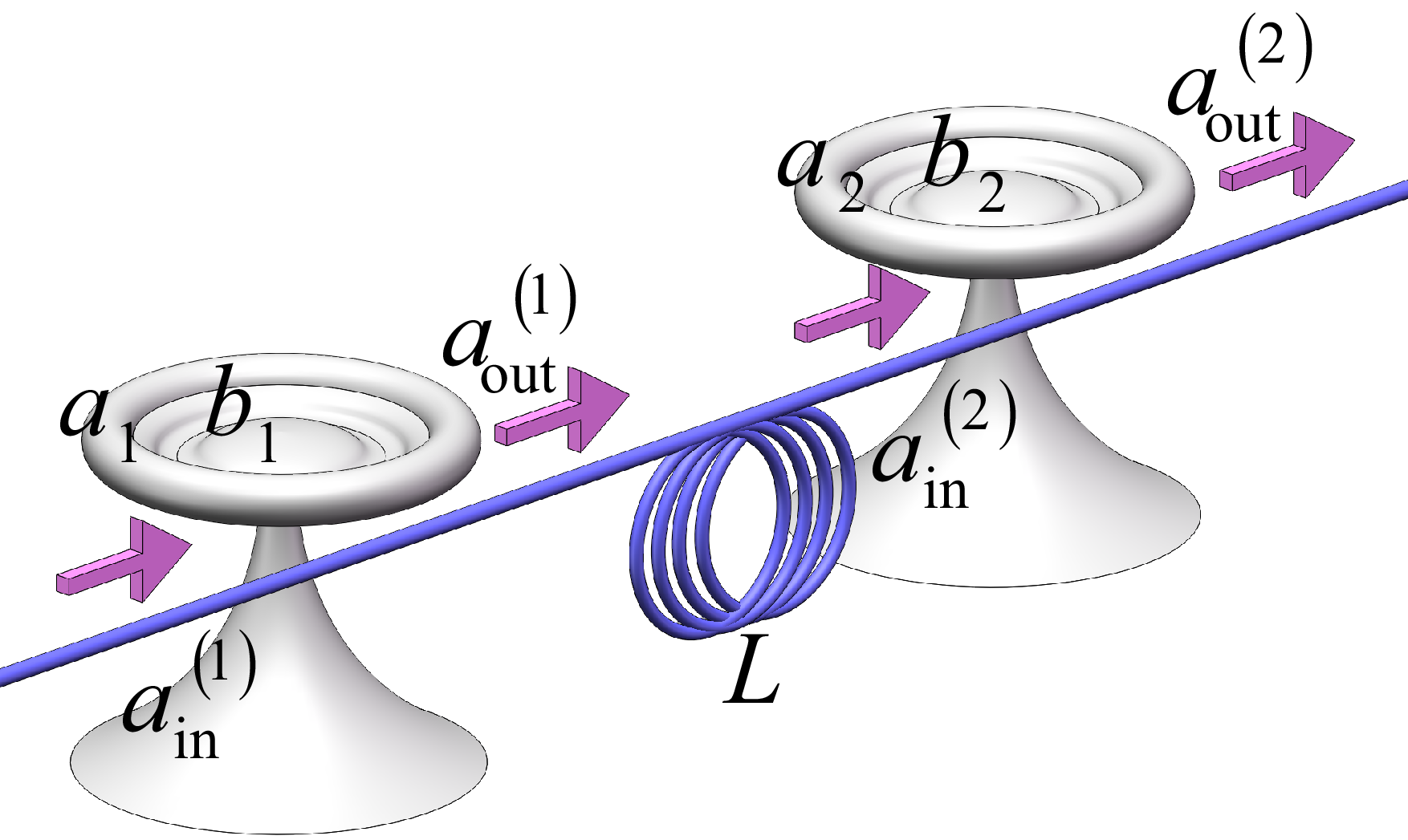}
\par\end{centering}

\protect\caption{(Color online) Schematic setup of the unidirectionally cascaded systems
of two toroid optical microcavities, coupled through a unidirectional
fiber with light. The distance between the two OMSs is denoted as
$L$. \label{fig:Setup of two OMSs}}
\end{figure}

The unidirectionally cascaded synchronization scheme consists of two
toroid optical microcavities \cite{armani2003ultra} with small mechanical
frequency mismatch. Both toroids are cascaded coupled with the optical
fiber, as shown in Fig.$\,$\ref{fig:Setup of two OMSs}. The input
laser in the fiber is coupled to the traveling optical whispering
gallery modes in the former toroid, and the transmitted light is coupled
to the following toroid. Each toroid also supports low loss mechanical
breath vibration mode \cite{PhysRevLett.97.243905}, thus enables
optomechanical coupling. In our model, it's assumed that there is
no laser input in the reversal direction, so the optical coupling
between cascaded toroids are unidirectional. We would expect that
light could carry the vibration information from the first optomechanical
system (OMS-1) to the second optomechanical system (OMS-2), and thus
enable the unidirectional synchronization.

The Hamiltonian of the individual OMS-$j$ ($j=1,2$) is
\begin{equation}
H_{j}=\hbar\omega_{cj}a_{j}^{\dagger}a_{j}+\hbar\omega_{mj}b_{j}^{\dagger}b_{j}-\hbar g_{j}a_{j}^{\dagger}a_{j}\left(b_{j}+b_{j}^{\dagger}\right),\label{eq:System Hamiltonian}
\end{equation}
where $a_{j}^{\dagger}$($b_{j}^{\dagger}$) and $a_{j}$($b_{j}$)
are the optical (mechanical) creation and annihilation operators,
frequencies of optical and mechanical mode are denoted as $\omega_{cj}$
and $\omega_{mj}$ respectively. The last term describes the dispersive
coupling of optical mode and mechanical mode, where $g_{j}$ is the
vacuum optomechanical coupling rate.

The dynamics of the unidirectionally cascaded OMSs are determined
by the quantum Langevin equations, $\frac{\partial O}{\partial t}=\frac{1}{i\hbar}\left[O,H\right]+N-H_{diss}$,
where $O$ is an arbitrary system operator, $N$ and $H_{diss}$ represent
the environment noises and the system dissipation respectively. In
the semiclassical cases, the mean values of the environment noises
vanish, thus the equations of motion are as follows,

\begin{eqnarray}
\dot{a}_{j} & = & \frac{1}{i\hbar}\left[a_{j},H_{j}\right]-\frac{\kappa_{j}}{2}a_{j}+\sqrt{\kappa_{exj}}a_{in}^{\left(j\right)},\nonumber \\
\dot{b}_{j} & = & \frac{1}{i\hbar}\left[b_{j},H_{j}\right]-\frac{\gamma_{mj}}{2}b_{j},\label{eq:MEs of aj and bj}
\end{eqnarray}
where $j=1,2$. $\kappa_{j}$ and $\kappa_{exj}$ are the total and
external optical decay rates, respectively. $\gamma_{mj}$ is the
mechanical damping rate. $a_{in}^{\left(j\right)}$ represents the
injected driving field.

Based on the properties of cascaded systems \cite{PhysRevLett.70.2269,PhysRevLett.78.3221,PhysRevA.87.022318},
\begin{equation}
a_{in}^{\left(2\right)}\left(t\right)=\eta_{12}a_{out}^{\left(1\right)}\left(t-\tau\right),
\end{equation}
where $\eta_{12}=\sqrt{\eta_{P}}$ and $\eta_{P}$ is the power transmittance,
$a_{out}^{\left(1\right)}\left(t-\tau\right)$ represents the output
field of OMS-1, $\tau$ is the required time for light transmitting
from OMS-1 to OMS-2. In the case of unidirectionally cascaded systems,
only one direction for transmission is allowed. Thus, without loss
of universality, we let $\tau\rightarrow0_{+}$. Based on the input
and output theory of optical cavities \cite{walls2007quantum},
\begin{equation}
a_{out}^{\left(1\right)}\left(t\right)=a_{in}^{\left(1\right)}\left(t\right)-\sqrt{\kappa_{ex1}}a_{1}.
\end{equation}
Assuming $a_{in}^{\left(1\right)}\left(t\right)=E_{in}e^{-i\omega_{L}t}$,
where $E_{in}$ and $\omega_{L}$ represent the strength and frequency
of the driving optical field, respectively. In the rotating frame
with the driving frequency $\omega_{L}$, define $\tilde{a}_{j}=a_{j}e^{i\omega_{L}t}$
($j=1,2$), then based on Eq.$\,$(\ref{eq:MEs of aj and bj}), the
optical and mechanical modes satisfy
\begin{eqnarray}
\dot{\tilde{a}}_{1} & = & -(i\varDelta_{1}+\frac{\kappa_{1}}{2})\tilde{a}_{1}+ig_{1}\tilde{a}_{1}\left(b_{1}+b_{1}^{\dagger}\right)+E,\label{eq:a1TildeDot}\\
\dot{\tilde{a}}_{2} & = & -(i\varDelta_{2}+\frac{\kappa_{2}}{2})\tilde{a}_{2}+ig_{2}\tilde{a}_{2}\left(b_{2}+b_{2}^{\dagger}\right)\nonumber \\
 &  & +\eta_{12}\sqrt{\kappa_{ex2}}\left(E/\sqrt{\kappa_{ex1}}-\sqrt{\kappa_{ex1}}\tilde{a}_{1}\right),\label{eq:a2TildeDot}\\
\dot{b}_{j} & = & -i\omega_{mj}b_{j}+ig_{j}\tilde{a}_{j}^{\dagger}\tilde{a}_{j}-\frac{\gamma_{mj}}{2}b_{j},\thinspace\thinspace j=1,2,\label{eq:bjDot}
\end{eqnarray}
where $\varDelta_{j}=\omega_{cj}-\omega_{L}\thinspace(j=1,2)$ is
the driving field detuning. $E=\sqrt{\kappa_{ex1}}E_{in}$ is the
effective optical driving strength of OMS-1.

In addition, the equations of motion can also be derived consistently
from the master equation \cite{gardiner2004quantum}, indicating the
time evolution of density matrix $\rho$,

\begin{eqnarray}
\dot{\rho} & = & \frac{1}{i\hbar}\left[H_{1}+H_{2},\rho\right]+\kappa_{1}\mathcal{L}\left[a_{1}\right]\rho+\kappa_{2}\mathcal{L}\left[a_{2}\right]\rho\nonumber \\
 &  & +\gamma_{m1}\mathcal{L}\left[b_{1}\right]\rho+\gamma_{m2}\mathcal{L}\left[a_{2}\right]\rho\nonumber \\
 &  & +\sqrt{\kappa_{ex1}\kappa_{ex2}}\left\{ \mathcal{L}\left[a_{1}+a_{2}\right]\rho+\frac{1}{2}\left[a_{1}^{\dagger}a_{2}-a_{2}^{\dagger}a_{1},\rho\right]\right\} \nonumber \\
 &  & +\left[a_{in}^{\left(1\right)}\left(\sqrt{\kappa_{ex1}}a_{1}^{\dagger}+\sqrt{\kappa_{ex2}}a_{2}^{\dagger}\right)-h.c,\rho\right],\label{eq:MasterEquation}
\end{eqnarray}
where $\mathcal{L}\left[o\right]\rho=o\rho o^{\dagger}-\frac{1}{2}\left(o^{\dagger}o\rho+\rho o^{\dagger}o\right)$
is the Lindblad superoperator. And the coupling term in the master
equation consists of a damping term $\mathcal{L}\left[a_{1}+a_{2}\right]\rho$
and a commutator $\frac{1}{2}\left[a_{1}^{\dagger}a_{2}-a_{2}^{\dagger}a_{1},\rho\right]$,
which indicates the system's unidirectionality.

From equations of motion {[}Eqs.$\,$(\ref{eq:a1TildeDot}) and (\ref{eq:a2TildeDot}){]},
the output of OMS-1 drives the optical mode of OMS-2. In contrast,
the output of OMS-2 has no effects on OMS-1. Due to the nonlinear
interaction between optical mode and mechanical mode, such as $g_{1}\tilde{a}_{1}\left(b_{1}+b_{1}^{\dagger}\right)$
in Eq.\,(\ref{eq:a1TildeDot}), the output optical field from OMS-1
can modify the behavior of the mechanical resonator in OMS-2, and
may lead to the synchronization. The dynamics of the system is significantly
different from previously studied bidirectionally coupled OMSs, where
the mutual coupling can induce the synchronization.

\section{Unidirectional synchronization}

Since the nonlinear optomechanical interaction is crucial in the synchronization,
we don't apply linear approximations to solve the equations of motion.
The full dynamics of unidirectionally cascaded systems are simulated
for long evolution time numerically. For the convenient to illustrate
the synchronization, the optical and mechanical operators are re-written
as

\begin{equation}
\begin{array}{cccccc}
Q_{j} & = & \left(\tilde{a}_{j}+\tilde{a}_{j}^{\dagger}\right)/2, & P_{j} & = & -i\left(\tilde{a}_{j}-\tilde{a}_{j}^{\dagger}\right)/2,\\
q_{j} & = & \left(b_{j}+b_{j}^{\dagger}\right)/\sqrt{2}, & p_{j} & = & -i\left(b_{j}-b_{j}^{\dagger}\right)/\sqrt{2},
\end{array}
\end{equation}
where $j=1,2$. And the corresponding equations for quadratures of
optical fields $Q_{j},\thinspace P_{j}$ and mechanical displacement
$q_{j}$ and momentum $p_{j}$ read
\begin{eqnarray}
\dot{Q}_{1} & = & \left(\varDelta_{1}-G_{1}q_{1}\right)P_{1}-\frac{\kappa_{1}}{2}Q_{1}+E,\nonumber \\
\dot{P}_{1} & = & -\left(\varDelta_{1}-G_{1}q_{1}\right)Q_{1}-\frac{\kappa_{1}}{2}P_{1},\nonumber \\
\dot{Q}_{2} & = & \left(\varDelta_{2}-G_{2}q_{2}\right)P_{2}-\frac{\kappa_{2}}{2}Q_{2}+\nonumber \\
 &  & \sqrt{\kappa_{ex2}}\left(E/\sqrt{\kappa_{ex1}}-\sqrt{\kappa_{ex1}}Q_{1}\right),\nonumber \\
\dot{P}_{2} & = & -\left(\varDelta_{2}-G_{2}q_{2}\right)Q_{2}-\frac{\kappa_{2}}{2}P_{2}-\sqrt{\kappa_{ex2}\kappa_{ex1}}P_{1},\nonumber \\
\dot{q}_{j} & = & \omega_{mj}p_{j},\nonumber \\
\dot{p}_{j} & = & -\omega_{mj}q_{j}-\gamma_{mj}p_{j}+G_{j}\left(\tilde{a}_{jr}^{2}+\tilde{a}_{ji}^{2}\right),\label{eq:real ME of two OMSs}
\end{eqnarray}
where $G_{j}=\sqrt{2}g_{j}$. The numerical simulation is performed
using the four-order Runge-Kutta algorithm. In the simulation, we
choose realistic values of the parameters \cite{li2015criterion,aspelmeyer2014cavityArticle}
and normalize them by $\omega_{m1}$: $\omega_{m1}=1,$ $\omega_{m2}=1.005,$
i.e., the intrinsic frequency of OMS-2 differs from that of OMS-1
with a mismatch of $5\permil\omega_{m1}$. $\varDelta_{1}=-\omega_{m1}$,
$\varDelta_{2}=-\omega_{m2}$, i.e., the driving laser is blue detuned,
which guarantees that OMS-1 will evolve into self-sustained oscillation
as long as the driving strength is strong enough. The other parameters
are $G_{1}=G_{2}=4\times10^{-3}$, $\kappa_{1}=\kappa_{2}=0.15,$
$\kappa_{ex1}=\frac{\kappa_{1}}{2}$, $\kappa_{ex2}=\frac{\kappa_{2}}{2}$,
$\gamma_{m1}=\gamma_{m2}=5\times10^{-3}$. In addition, the time scale
in the simulation becomes dimensionless and changes from $t$ into
$t'=\omega_{m1}t$ due to the normalization.

\begin{figure}[th]
\begin{centering}
\includegraphics[width=8.5cm]{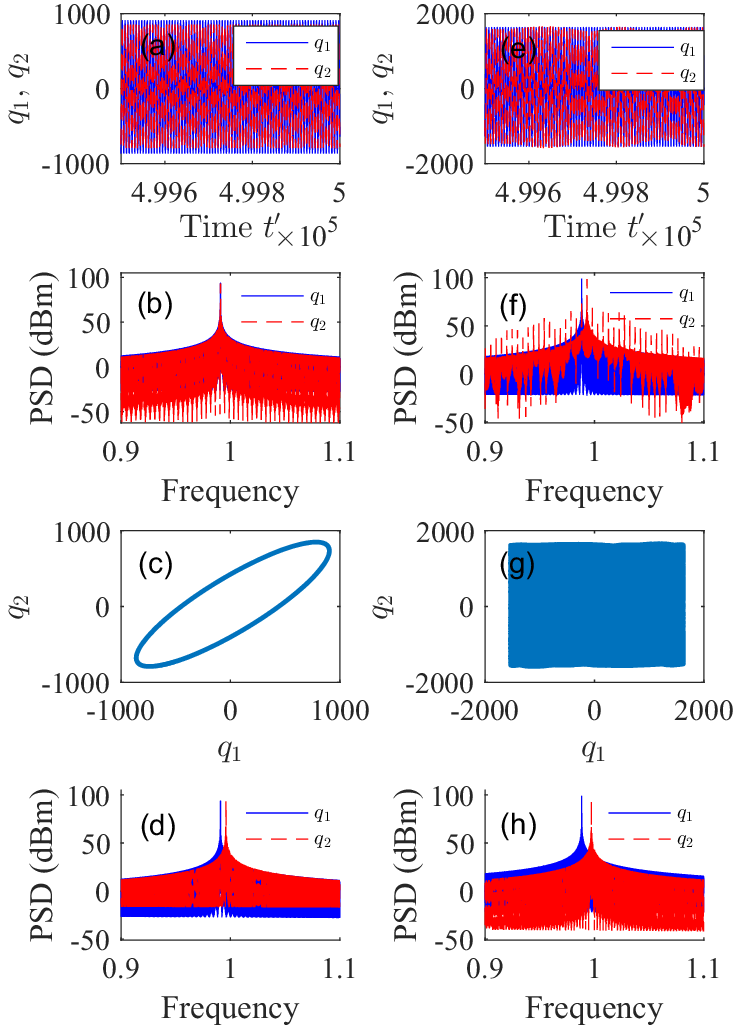}
\par\end{centering}

\protect\caption{(Color online) Numerical solutions of equations of motion in the unidirectionally
cascaded two OMSs scheme, for the parameters: $\omega_{m1}=1$, $\omega_{m2}=1.005$,
$\varDelta_{1}=-\omega_{m1}$, $\varDelta_{2}=-\omega_{m2}$, $G_{1}=G_{2}=4\times10^{-3},$
$\kappa_{1}=\kappa_{2}=0.15,$ $\kappa_{ex1}=\kappa_{ex2}=0.075$,
$\gamma_{m1}=\gamma_{m2}=5\times10^{-3}$. (a) and (e) Dynamical evolution.
(b) and (f) Power spectrum density (PSD) of the displacement operators
$q_{1}$ and $q_{2}$. (c) and (g) Phase diagram of $q_{1}$, $q_{2}$.
(d) and (h) Displacement PSD of each single OMS driven by a constant
amplitude optical field $E$. The left and right columns correspond
to $E=64$ and $E=100$. \label{fig:Unidirectional Syn, E=00003D64, E=00003D100}}
\end{figure}

Firstly, we study the general properties of lossless cascade coupling
between two OMSs. Figure \ref{fig:Unidirectional Syn, E=00003D64, E=00003D100}
shows typical behaviors of the OMSs for different parameters. Under
the effective driving of $E=64$, the dynamical evolution of mechanical
displacement for two OMSs are shown in Fig.$\,$\ref{fig:Unidirectional Syn, E=00003D64, E=00003D100}(a),
and the corresponding power spectrum density (PSD) and phase diagram
are shown in Fig.\,\ref{fig:Unidirectional Syn, E=00003D64, E=00003D100}(b)
and Fig.\,\ref{fig:Unidirectional Syn, E=00003D64, E=00003D100}(c),
respectively. Although the intrinsic mechanical frequencies are different
by $5\permil$, the eventual oscillation frequencies are the same
$\omega'_{m1}=\omega'_{m2}=0.990932$. The regular orbit in phase
diagram confirms that the oscillation periods are exactly the same
and their phase difference is constant. To exclude the possible coincidence
that the self-oscillation frequencies of two OMSs under external optical
driving are the same, we also plot the PSD for OMSs individually driven
by the external laser in Fig.$\,$\ref{fig:Unidirectional Syn, E=00003D64, E=00003D100}(d).
The steady state oscillation frequencies are $\omega'_{m1}=0.990932$,
$\omega'_{m2}=0.995881$, which are different by about $5\permil$
similar to that of intrinsic frequencies. The results confirm that
$\omega'_{m1}$ are exactly the same as $\omega_{m1}$ and not affected
by the OMS-2, and the OMS-2 is synchronized to OMS-1 under the unidirectional
optical coupling.

With a further increase in the strength of the laser driving the OMSs,
the two OMSs are not guaranteed to be synchronized under the unidirectional
coupling. As shown in Figs.\,\ref{fig:Unidirectional Syn, E=00003D64, E=00003D100}(e)-\ref{fig:Unidirectional Syn, E=00003D64, E=00003D100}(h),
the OMSs are unsynchronized for $E=100$. From the PSD, the OMS-1
is still unaffected by the OMS-2, just shows a single peak self-oscillation
behavior. However, the PSD of OMS-2 {[}Fig.$\,$\ref{fig:Unidirectional Syn, E=00003D64, E=00003D100}(f){]}
shows multiple peaks when driven by the output from OMS-1. The frequency
locations of those peaks show equal distances. This can be interpreted
as the dynamics of OMS-2 can still be greatly affected by OMS-1 for
large laser driving, but nonlinear effect generates the frequency
mixing of two systems instead of synchronization, which is a typical
feature of the well-known nonlinear periodic pulling \cite{PhysRevA.44.6877,PhysRevE.52.4316,PhysRevE.84.016405}.

It is quite straightforward that there is also a threshold for nonlinear
optomechanical interaction to make synchronization happen. The above
results also indicate that the synchronization effect can only dominate
the other nonlinear effects in certain driving laser amplitudes. For
example, very strong nonlinear effect will induce multi-stable and
even chaotic dynamics. Therefore, we further study the final frequencies
$\omega'_{m1}$, $\omega'_{m2}$ as functions of the effective driving
strength $E$. For each $E$, we try 10 sets of different random initial
values of the system to test the sensitivity of the synchronization
to initial conditions.

\begin{figure}[th]
\begin{centering}
\includegraphics[width=8.5cm]{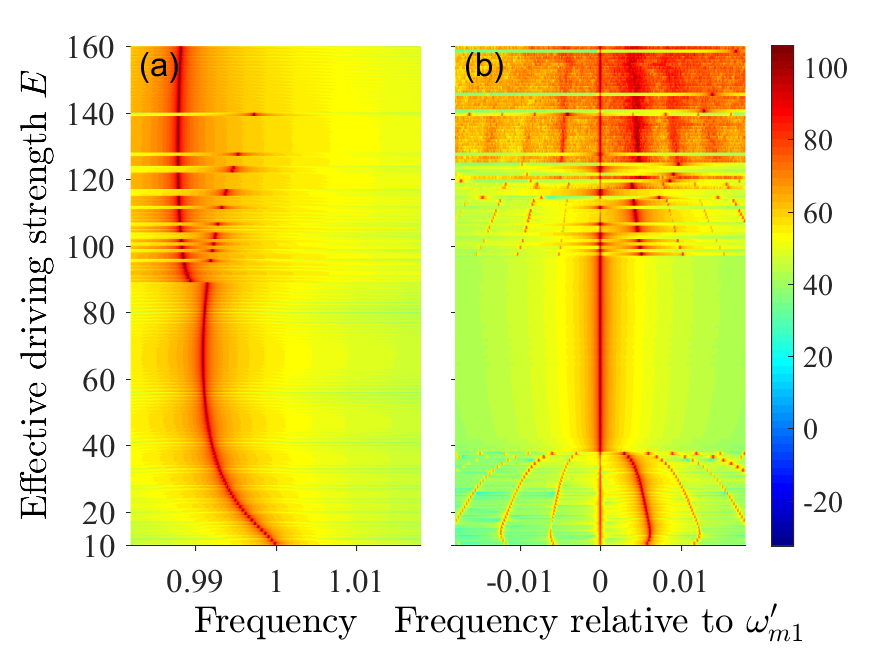}
\par\end{centering}

\protect\caption{(Color online) (a) The PSD of $q_{1}$ as a function of the effective
driving strength $E$, with a sole frequency peak denoted as $\omega'_{m1}$.
(b) The PSD of $q_{2}$ relative to $\omega'_{m1}$ as a function
of the effective driving strength $E$. The maximum frequency component
in the PSD of $q_{2}$ without the frequency shift is denoted as $\omega'_{m2}$.
The color scaled in the color bar indicates the power values in the
PSD. The other simulation parameters are the same as those in Fig.\,\ref{fig:Unidirectional Syn, E=00003D64, E=00003D100}.
\label{fig:E influence on Syn} }
\end{figure}

In Fig.\,\ref{fig:E influence on Syn}, the PSD of $q_{1}$ and the
PSD of $q_{2}$ but shifted in respect to the spectrum of $q_{1}$
are plotted. The results reveals different dynamical regimes for unidirectionally
coupled OMSs: (1) Weak nonlinear effect, $E\in\left[10,37\right]$.
Below the threshold of about $E\approx37$, the two OMSs are unsynchronized
$\omega'_{m2}\neq\omega'_{m1}$. However, the OMS-2 are affected by
the mechanical oscillation in OMS-1, thus a series of sidebands appear
in the PSD of $q_{2}$. (2) Synchronization, $E\in\left[38,96\right]$.
For moderate driving, the two OMSs are synchronized with a sole peak
in the PSD of $q_{2}$ and $\omega'_{m2}=\omega'_{m1}$. (3) Multi-stable
and chaotic regime, $E\in\left[97,160\right]$. For very strong driving,
the two OMSs are unsynchronized. There are multiple possible self-oscillation
frequencies of OMS-1. For certain OMS-1 oscillation frequency, the
OMS-2 can still be synchronized. While, for other frequencies, the
OMS-2 exhibits very complex dynamics, including synchronization, frequency
mixing and multi-stable dynamics simultaneously.

Actually, due to the unidirectionality, OMS-1 is independent from
OMS-2 and thus can be fully theoretically solved using the single
OMS theory \cite{PhysRevLett.96.103901} and the output field is modulated
by the mechanical vibration. The observed synchronization and periodic
pulling phenomena of OMS-2 originate from the modulated laser driving
on OMS-2. Similar effects have been demonstrated with the injection-locking
\cite{Adler1946Locking,Razavi2004InjectionLocking,PhysRevLett.105.013004}
of an OMS \cite{shlomi2015synchronization,hossein2008observation,shah2015master},
where the input laser of the OMS is partially modulated by a single
tone RF signal using an electro-optic modulator.

Previous studies show that synchronization occurs only when the driving
RF frequency is very close to the intrinsic oscillation frequency
\cite{PhysRevE.52.4316}. Inspirited by this, the final relative frequency
difference $\left(\omega'_{m2}-\omega'_{m1}\right)/\omega'_{m1}$
as a function of the intrinsic frequency mismatch $\left(\omega_{m2}-\omega_{m1}\right)/\omega_{m1}$
is plotted in Fig.\,\ref{fig:wm2 influence on Syn, E=00003D40, E=00003D64}.
When $E=64$ ($E=40$), there is a synchronization region of $\omega_{m2}\in\left[1-9.2\permil,1+8.2\permil\right]\omega_{m1}$
($\omega_{m2}\in\left[1-8\permil,1+5.6\permil\right]\omega_{m1}$),
represented by the red line (the blue line). We find that, the width
of synchronization region is similar to the mechanical damping rate
$5\permil$ \cite{hossein2008observation}, and the increase of driving
strength does enlarge the width of synchronization region, by comparing
the results of $E=64$ and that of $E=40$.

\begin{figure}[th]
\begin{centering}
\includegraphics[width=8cm]{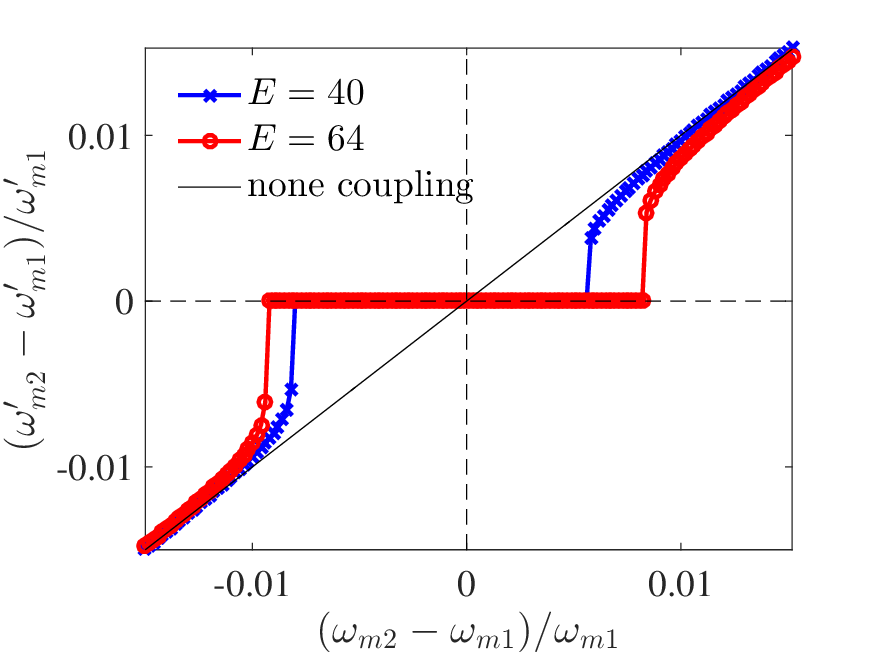}
\par\end{centering}

\protect\caption{(Color online) The final relative frequency difference between the
maximum frequency components of two OMSs $\left(\omega'_{m2}-\omega'_{m1}\right)/\omega'_{m1}$
vs the relative intrinsic frequency mismatch $\left(\omega_{m2}-\omega_{m1}\right)/\omega_{m1}$.
Blue solid line marked with 'x': $E=40$. Red solid line marked with
'o': $E=64$. Black solid line: uncoupled free-running case. \label{fig:wm2 influence on Syn, E=00003D40, E=00003D64}}
\end{figure}

\section{Long distance unidirectional synchronization with fiber-loss}

The unidirectional coupling is very potential for future long distance
synchronization, since the OMSs are directly coupled through the optical
connections, without extra optical-to-electronic or reversal conversion
processes. In addition, the unidirectional coupling also greatly reduces
the complexity of experiments. To testify the potential for long distance
synchronization, we take the practical fiber attenuation loss into
our model. Take 1550nm light as an example, the propagation loss rate
is $\alpha=0.2\thinspace{\rm dB}/{\rm km}$, the power transmittance
$\eta_{P}=10^{-\alpha L/10}$.

\begin{figure}[th]
\begin{centering}
\includegraphics[width=8.5cm]{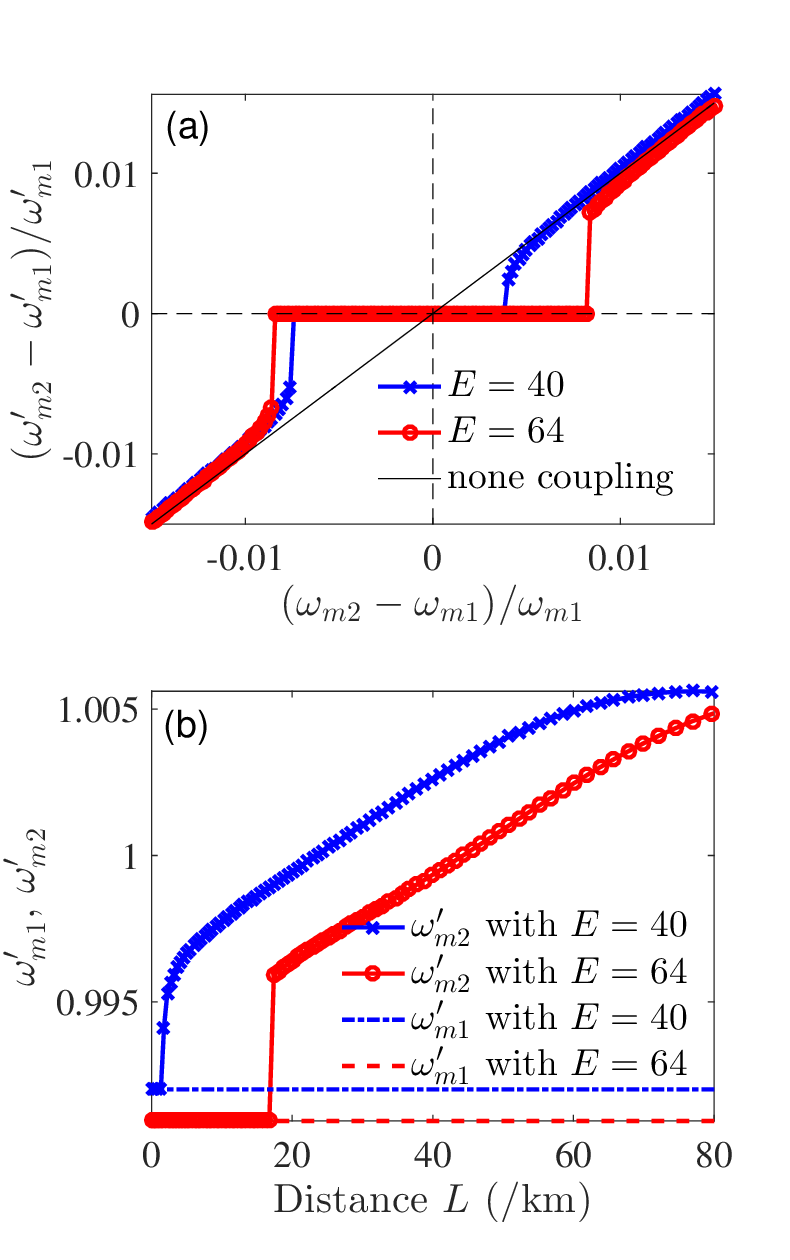}
\par\end{centering}

\protect\caption{(Color online) Considering fiber-loss, (a) the final relative frequency
difference between the maximum frequency components of the two OMSs
$\left(\omega'_{m2}-\omega'_{m1}\right)/\omega'_{m1}$ vs the relative
intrinsic frequency mismatch $\left(\omega_{m2}-\omega_{m1}\right)/\omega_{m1}$
with $L=4.6\thinspace{\rm km}$, $\eta_{12}=0.9$. The cases corresponding
to the lines are the same as those in Fig.\,\ref{fig:wm2 influence on Syn, E=00003D40, E=00003D64}.
(b) The final maximum frequency components of the two OMSs $\omega'_{m1}$,
$\omega'_{m2}$ vs distance $L$. Blue solid line marked with 'x':
$\omega'_{m2}$ with $E=40$. Red solid line marked with 'o': $\omega'_{m2}$
with $E=64$. Blue dash-dotted line: $\omega'_{m1}$ with $E=40$.
Red dashed line: $\omega'_{m1}$ with $E=64$. \label{fig:wm2 influence on Syn, E=00003D40, E=00003D64, L=00003D6.4}}
\end{figure}

First, take $L=4.6\,\mathrm{km}$, i.e., $\eta_{12}=\sqrt{\eta_{P}}=0.9$,
synchronization region of $\omega_{m2}\in\left[1-8.4\permil,1+8.2\permil\right]\omega_{m1}$
($\omega_{m2}\in\left[1-7.4\permil,1+3.8\permil\right]\omega_{m1}$)
is revealed for $E=64$ ($E=40$) {[}Fig.\,\ref{fig:wm2 influence on Syn, E=00003D40, E=00003D64, L=00003D6.4}(a){]}.
Compared to the result without fiber loss $\eta_{P}=1.0$ in Fig.\,\ref{fig:wm2 influence on Syn, E=00003D40, E=00003D64},
the parameter region has shrunk. Then, we explore whether the systems
are synchronized or not for $L$ varying from 0 to 80km, while fixing
the intrinsic mechanical frequencies $\omega_{m1}=1,$ $\omega_{m2}=1.005$.
As plotted in Fig.\,\ref{fig:wm2 influence on Syn, E=00003D40, E=00003D64, L=00003D6.4}(b),
we can see for each $E$ there exists a critical distance $L_{cri}$,
over which the state of two OMSs changes from synchronization into
unsynchronization. The critical distance for $E=40$ and $E=64$ are
as long as 1.3 km and 16.7 km, respectively, which verifies the capability
of our scheme to realize long-distance unidirectional synchronization.

\section{Unidirectional synchronization of three OMSs}

\begin{figure}[th]
\begin{centering}
\includegraphics[width=8.5cm]{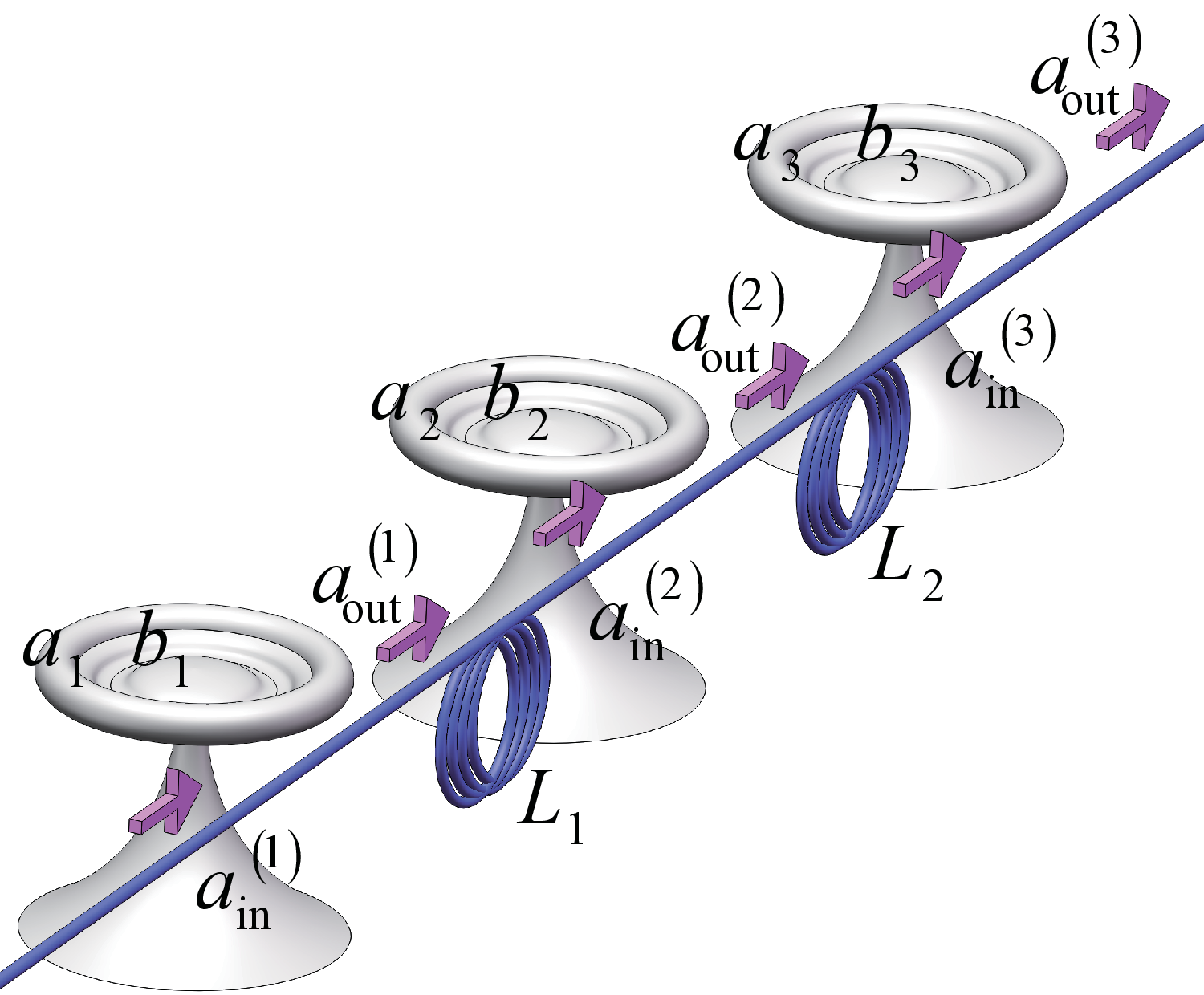}
\par\end{centering}

\protect\caption{(Color online) Schematic setup of the unidirectionally cascaded synchronization
scheme consists of three OMSs. The distance between the first (last)
two OMSs is denoted as $L_{1}$ ($L_{2}$). \label{fig:Setup of three OMSs}}
\end{figure}

Now, we further study the generalized unidirectional synchronization
of a cascaded OMSs array. From the results of two OMSs, the synchronization
is possible for additional OMSs following OMS-2, as long as the driving
laser intensity on them is moderate and contains the components of
modulation from the mechanical vibrations. As an example, the unidirectional
synchronization scheme of three dissimilar OMSs (microtoroids) with
different intrinsic frequencies are cascaded using a unidirectional
fiber, as shown in Fig.\,\ref{fig:Setup of three OMSs}.

Let $\eta_{23}=a_{in}^{\left(3\right)}/a_{out}^{\left(2\right)}$,
$\eta_{12}=a_{in}^{\left(2\right)}/a_{out}^{\left(1\right)}$, the
set of equations of motion can be obtained, where $\dot{\tilde{a}}_{1}$,
$\dot{\tilde{a}}_{2}$, $\dot{b}_{1}$ and $\dot{b}_{2}$ are the
same as Eqs.(\ref{eq:a1TildeDot},\ref{eq:a2TildeDot},\ref{eq:bjDot})
due to the unidirectionality, and $\dot{\tilde{a}}_{3}$ and $\dot{b}_{3}$
are in the following form,

\begin{eqnarray}
\dot{\tilde{a}}_{3} & = & -i\left(\varDelta_{3}+\frac{\kappa_{3}}{2}\right)\tilde{a}_{3}+ig_{3}\tilde{a}_{3}\left(b_{3}+b_{3}^{\dagger}\right)\nonumber \\
 &  & +\eta_{23}\eta_{12}\sqrt{\kappa_{ex3}}\left(E/\sqrt{\kappa_{ex1}}-\sqrt{\kappa_{ex1}}\tilde{a}_{1}\right)\nonumber \\
 &  & -\eta_{23}\sqrt{\kappa_{ex3}\kappa_{ex2}}\tilde{a}_{2},\nonumber \\
\dot{b}_{3} & = & -i\omega_{m3}b_{3}+ig_{3}\tilde{a}_{3}^{\dagger}\tilde{a}_{3}-\frac{\gamma_{m3}}{2}b_{3}.\label{eq:MEs of three OMSs}
\end{eqnarray}

Following the similar procedure of numerical simulation for two OMSs,
the dynamics of the three OMSs are solved. Shown in Fig.\,\ref{fig:PSD and wm3 influence on Syn, three OMOs}(a)
(Fig.\,\ref{fig:PSD and wm3 influence on Syn, three OMOs}(b)) the
PSDs of $q_{1}$, $q_{2}$ and $q_{3}$ under a set of parameters:
$\omega_{m2}=0.995$, $\omega_{m3}=1.005$ ($\omega_{m2}=1.005$,
$\omega_{m3}=1.010$) are plotted. The other simulation parameters
are $E=64$, $\omega_{m1}=1$, $\eta_{12}=\eta_{23}=1$, $\varDelta_{1}=\varDelta_{2}=\varDelta_{3}=-\omega_{m1}$,
$G_{1}=G_{2}=G_{3}=0.004,$ $\kappa_{1}=\kappa_{2}=\kappa_{3}=0.15,$
$\kappa_{ex1}=\kappa_{ex2}=\kappa_{ex3}=0.075$, $\gamma_{m1}=\gamma_{m2}=\gamma_{m3}=0.005$.
In Fig.\,\ref{fig:PSD and wm3 influence on Syn, three OMOs}(a),
the three OMSs are all synchronized, while in Fig.\,\ref{fig:PSD and wm3 influence on Syn, three OMOs}(b)
OMS-2 other than OMS-3 is synchronized to OMS-1. i.e., the three OMSs
are partially synchronized.

\begin{figure}[th]
\begin{centering}
\includegraphics[width=8.5cm]{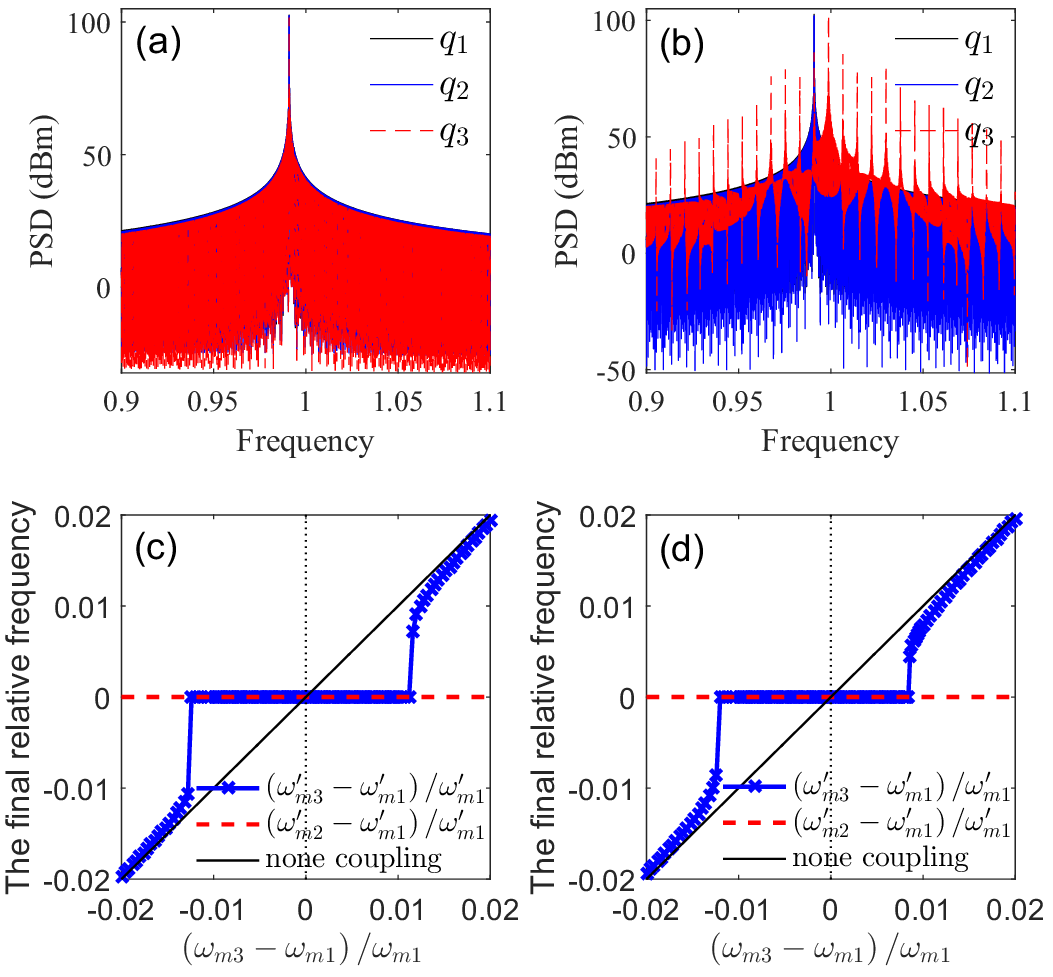}
\par\end{centering}

\protect\caption{(Color online) Numerical solutions of the equations of motion in the
unidirectionally cascaded three OMSs scheme. (a) and (b) The PSDs
of $q_{1}$, $q_{2}$ and $q_{3}$. The parameters are: (a) $\omega_{m2}=0.995$,
$\omega_{m3}=1.005$, (b) $\omega_{m2}=1.005$, $\omega_{m3}=1.010$.
(c) and (d) The synchronization region of $\omega_{m3}$ relative
to $\omega_{m1}$. The parameters are: (c) $\omega_{m1}=1$, $\omega_{m2}=0.995$,
(d) $\omega_{m1}=1$, $\omega_{m2}=1.005$. \label{fig:PSD and wm3 influence on Syn, three OMOs}}
\end{figure}

In addition, for a fixed effective driving strength $E=64$ and $\omega_{m1}=1$,
when $\omega_{m2}=0.995$ ($\omega_{m2}=1.005$), the synchronization
region of $\omega_{m3}\in\left[1-12.4\permil,1+11.2\permil\right]\omega_{m1}$
($\omega_{m3}\in\left[1-12\permil,1+8.6\permil\right]\omega_{m1}$)
can be obtained by traversing $\omega_{m3}\in\left[0.98,1.02\right]$,
as shown in Fig.\,\ref{fig:PSD and wm3 influence on Syn, three OMOs}(c)
(Fig.\,\ref{fig:PSD and wm3 influence on Syn, three OMOs}(d)). Comparing
the synchronization region of the three OMSs scheme with that of the
two OMSs scheme {[}Fig.\,\ref{fig:wm2 influence on Syn, E=00003D40, E=00003D64}{]},
we find that the synchronization region has expanded to some extent
at certain values of $\omega_{m2}$. Thus, the synchronization of
three OMSs is available, which verifies the feasibility of synchronization
of an array of more than 2 cascaded OMSs.

\section{\textit{\emph{Conclusion}}}

We have demonstrated the synchronization of optomechanical systems
by all-optical method, where the systems are coupled through light
propagating unidirectionally in the fiber. For two OMSs with fixed
mechanical frequency mismatch, synchronization can be tuned on or
off through tuning the optical driving strength. For a fixed driving
strength, there exists a region of mechanical frequency mismatch that
allows for the synchronization. And in the practical cases, the synchronization
can still be achieved for distance over 10 km, while the synchronization
region shrinks due to the light attenuates when travel over long distances.
Unidirectional synchronization of three OMSs is also obtained, as
well. The all-optical feature, high controllability, wide synchronization
region, long synchronization distance, and novel scalability of our
scheme are appealing and can be useful for many applications, such
as the construction of complex synchronization OMSs networks \cite{zhang2015synchronization}.
We expected that the scheme also works for other optomechanical interactions,
such as quadratic \cite{lee2015multimode} , dissipative \cite{PhysRevLett.103.223901,PhysRevA.84.032317}
and Brillouin \cite{dong2015brillouin,kim2015non} optomechanical
interactions.

\section*{Acknowledgments}

We thank C. H. Dong, C. S. Yang, Z Shen and Z. H. Zhou for useful
discussions. This work was funded by the National Basic Research Program
of China (Grant No. 2013CB338002 and No. 2011CBA00200), National Natural
Science Foundation of China (Grant No. 11074244 and No. 61502526).

\bibliographystyle{apsrev4-1}

%

\end{document}